\documentclass[twocolumn,prb,aps]{revtex4}
\usepackage{epsfig}
\begin{document}

\title{Structure and dynamics of the interface between a 
binary hard-sphere crystal of NaCl type and its coexisting binary fluid}

\author{Rachel Sibug-Aga and Brian B. Laird\footnote{Author to whom correspondence should be
addressed.}}
\affiliation{Department of Chemistry, University of Kansas, 
         Lawrence, Kansas 66045, USA}

\date{\today}

\begin{abstract}
Molecular dynamics simulations are performed to study the [100] and [111] 
orientations of the crystal-melt interface between an ordered two-component hard sphere 
with a NaCl structure and its coexisting binary hard-sphere fluid.
The diameter ratio of the two types of hard spheres making up the mixture is taken
to be $\alpha=0.414$.  This work complements our earlier interface simulations
[J. Chem. Phys. {\bf 116}, 3410] for the same diameter ratio at lower pressures where
the smaller component is immiscible in the solid and the fluid mixture coexists 
with a pure FCC crystal of large particles.
Density profiles and diffusion coefficient profiles are presented for the AB 
interfacial system.  We find that for this system, the transition from 
crystal-like to fluid-like behavior of both the density and diffusion constant
profiles occurs over a narrower region than that seen in our previous studies 
[J. Chem. Phys.  {\bf 116}, 3410] of the FCC/binary fluid system.
But similar to what was found in the FCC/binary fluid interface the transition region
for the large particle diffusion constant is shifted about 1.0$\sigma_A$ toward
the fluid phase relative to that for the small particles.
\end{abstract}
\maketitle

\section{Introduction}

The kinetics of crystal growth and nucleation from the melt is
highly dependent upon the structure, dynamics and thermodynamics
of the crystal-melt interface\cite{Tiller91}. Given the 
difficulties in obtaining unambiguous information from experiments,
most of what is currently known about the microscopic phenomenology of such
interfaces is obtained via computer simulation\cite{Laird98}. 
Although most simulation studies have focused on single component 
systems\cite{Broughton86b,Broughton86c,Karim88,Laird89c,Davidchack98,Hayward01,Hoyt01}, 
there has been recent interest in multicomponent interfaces\cite{Davidchack96,Davidchack99,
Sibug-Aga02,Hoyt02}. All of these studies have involved  crystal phases that are either
disordered or pure face-centered cubic (FCC) lattices. In this work we
present results for the structure and dynamics of the interface between
an {\it ordered} two-component hard-sphere crystal with a  sodium chloride (NaCl) structure
and a binary hard-sphere fluid. Such a system can be viewed either as as a prototype to
understand the interface between inter-metallic compounds and their coexisting fluid phases 
or as a model two-component colloidal dispersion. 

The hard-sphere interaction was chosen for this study since it is an 
important reference model for the study of simple liquids~\cite{Hansen86}
and liquid mixtures~\cite{Young93}. This is especially true with regard to 
phenomena associated with the freezing transition. For example, 
it has been recently shown that the interfacial free energy of 
close-packed metals can be described with quantitative accuracy
using a hard-sphere model~\cite{Laird01}.
In addition, recent phase boundary calculations have shown that binary hard spheres 
form a wide range of crystal structures depending on the 
ratio, $\alpha =\sigma_B/\sigma_A$, of the diameter of the small spheres (labeled B), 
$\sigma_B$ to that of the larger spheres (labeled A), $\sigma_A$.  A substitutionally disordered FCC 
crystal is the stable phase for  
$1.0>\alpha>0.85$ ~\cite{Kranendonk91a}  while for  $\alpha<0.85$, only 
ordered crystal structures are seen to be stable, including 
AB, AB$_2$ and AB$_{13}$ structures~\cite{Trizac97,Cottin95, 
Eldridge93a,Eldridge93b,Eldridge95}.  A detailed study of 
the disordered FCC crystal/melt interface for $\alpha=0.9$ has been 
recently reported~\cite{Davidchack99}.  

\begin{figure}[h!]
\vspace{0.2cm}
\centerline{
\epsfig{file=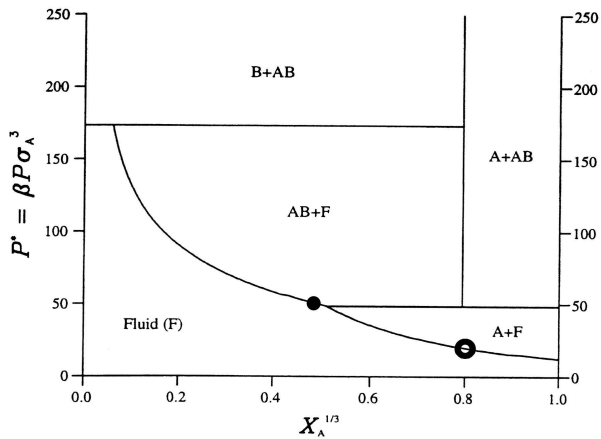,height=6.3cm,width=9.0cm,angle=0}}
\caption{\small Pressure-concentration phase diagram of the binary hard-sphere
system with $\alpha = 0.414$. [Reprinted from Ref.\cite{Trizac97} by permission
of the publisher, Taylor and Francis, Ltd (www.tandf.co.uk/journals)]. Note
that to make the phase coexistence lines easier to distinguish, the pressure
is plotted against the cube root of the large sphere mole fraction. The phase point
of the present study is shown as a filled circle. The open circle shows the 
conditions of our previous study\cite{Sibug-Aga02}.} 
\label{phasediag}
\vspace{0.2cm}
\end{figure}

In this work, we examine two-component hard-sphere mixtures with a considerably larger
size asymmetry of $\alpha=0.414$.  This size ratio is significant in the theory of binary
alloys in that it is the largest asymmetry in which the small spheres can be accommodated 
in the interstitials of a densest close-packed crystal of larger spheres.
The phase diagram for this value of $\alpha$  
has been determined as a function of pressure and mole fraction using MC and 
MD simulations by Trizac and coworkers\cite{Trizac97} and is reproduced 
in Fig.~\ref{phasediag}.  
At low pressures, the binary fluid coexists with
a pure FCC crystal of large spheres, whereas at higher pressures
(above $50 kT/\sigma_A^3$) the coexisting solid phase is an ordered 1:1 crystal
of the sodium chloride (NaCl) type.  Earlier cell theory calculations 
also predicted the stability of the NaCl at this diameter 
ratio~\cite{Cottin95}.  Other AB structures such as the CsCl and 
the zinc blende have been shown to be unstable at this diameter 
ratios~\cite{Cottin95,Eldridge95}.  Throughout the text we will be using AB to also refer 
to the NaCl structure. 

A detailed study of the low pressure pure FCC/binary fluid system has been 
recently reported~\cite{Sibug-Aga02}.  In that work, the coexistence between
the crystal and an approximately 1:1 binary mixture was examined, corresponding
to a pressure of $20.1 kT/\sigma_A^3$, which is approximately twice the coexistence
pressure ($11.55 kT/\sigma$) of the pure single component system.  
As the pressure is increased the mole fraction of large spheres in the fluid phase, $X^f_A$, decreases, and
at a pressure of about $50 kT/\sigma_A^3$ the fluid coexists with an AB crystal
of NaCl type (see Fig.~\ref{phasediag}).  In this work, we examine in detail
the microscopic structure and dynamics of the interface between the high pressure
AB crystal and its melt. To do this, we have chosen a point in the phase 
diagram with $X^f_A=0.097$ .  At this mole fraction, the fluid coexists with 
the NaCl crystal at a pressure of $53 kT/\sigma_A^3$.  Details of the simulation
methodology and interface equilibration procedure are outlined in the next section followed
by presentation of the results of the study in Section III and concluding remarks in section IV.

\section{Simulation Details}
Molecular-dynamics simulations are performed on a two-component system of 
hard spheres of differing diameters, $\sigma_A$ and $\sigma_B$, with $\sigma_A < \sigma_B$
(Type $A$ particles are assumed to be the larger spheres). The interaction potential
between two spheres is defined by the following pairwise interaction potential 
\begin{equation}
\phi_{ij}(r)=\left\{\begin{array} 
             {r@{\quad,\quad}l}          
             \infty & r\le\sigma_{ij}  \\ 0 & r>\sigma_{ij}
             \end{array} \right.,
\end{equation}
where $i,j \in \{A,B\}$, $r$ is the distance between the centers of the two interacting spheres, 
and $\sigma_{ij} = (\sigma_i + \sigma_j)/2$ is the distance of closest 
possible approach between two spheres with diameter $\sigma_i$ and $\sigma_j$, respectively.  
The system is completely defined by specifying the diameter ratio $\alpha \equiv \sigma_B/\sigma_A$,
the mole fraction of large particles, $x_A$ and the total number density, $\rho$. 
The total volume occupied by the hard spheres relative 
to the volume available to the system is given by the packing fraction, 
\begin{equation}
\eta=\frac{\pi \sigma_A^3}{6} \rho [x_A  + (1-x_A)\alpha^3],
\end{equation}
where $\rho = \rho_A + \rho_B = N/V$ is the total density, 
$x_A$ is the mole fraction of the larger species, and  
$\alpha = \sigma_B/\sigma_A$ is the diameter ratio.

The procedures for interface construction and equilibration of binary interfaces employed 
in this study are similar to those used in our earlier work\cite{Sibug-Aga02} on the low-pressure
coexistence in this system (single component FCC ($x_A=1$)/binary fluid mixture). 
Since the general interface preparation process is described at length in that work, only
those details specific to the current system are described here. The interested reader is 
encouraged to consult reference \cite{Sibug-Aga02} for a more complete description.

To construct an interface, fluid and crystal blocks are 
prepared separately using the calculated coexistence conditions.
At a pressure of 53 $kT/\sigma_A^3$, we independently 
determined the packing fractions of the coexisting crystal and
fluid to be $\eta^c = 0.683$ and $\eta^f = 0.490$, respectively 
In the preparation of a fluid block, it is a usual practice to initially 
position particles in a lattice at a density lower than coexistence.  
As the system is allowed to equilibrate the lattice melts, giving a 
fluid configuration, which is then compressed to the coexistence 
density.  For the system under study here this procedure is not feasible since 
the size asymmetry ($\alpha=0.414$) 
and the mole fraction of small particles ($X_B^f=0.903$) in the fluid system 
are both large.  (It should be noted that for this value of $\alpha$ the large sphere
volume is over 14 times that of the small particle.)
To construct the fluid phase we began with an FCC lattice 
of small particles at a number density equal to the desired total number density of the
fluid mixture.  A number of particles, corresponding to the target mole fraction
of large spheres, are then chosen at random from this lattice.
A molecular dynamics run is started from this initial configuration and the diameter of 
randomly chosen particles is periodically increased until the correct large sphere diameter 
is attained.  The amount of increase in the diameter at each stage depends 
on the maximum increase that is possible without creating particle overlap.
The preparation of the crystal is straightforward as the small particles easily
inserted into the interstitial sites of the large sphere FCC lattice. 

After equilibration of separate crystal and fluid systems,  they are  placed in contact
within the simulation box. Due to the periodic boundary conditions, two interfaces are formed.  
Note that packing fraction used in the preparation of the initial 
fluid block is slightly different from the predicted $\eta^f$   
because a gap of $1 \sigma_A$ is  placed between the crystal and fluid blocks
to avoid any initial overlap that may occur when the two blocks 
are combined.  The fluid particles are then allowed to move while the 
large particles are fixed so the initial gaps are then filled with the 
fluid particles.
The initial fluid packing fraction is adjusted until an unstressed 
bulk crystal is obtained when the two blocks are combined and equilibrated

\begin{table}[t]
\begin{minipage}{8cm}
\caption{Number of particles and dimensions of the simulation box}
\end{minipage}
\vspace{0.2cm}
\begin{tabular}{c@{\hspace{0.55cm}}c@{\hspace{0.55cm}}c@{\hspace{0.55cm}}
  c@{\hspace{0.55cm}}c@{\hspace{0.55cm}}c@{\hspace{0.55cm}}c@{\hspace{0.55cm}}}
\hline \hline
        & $N^c$ & $N^f$ & $L_x/\sigma_A$ & $L_y/\sigma_A$ & $L_z/\sigma_A$  \\
\hline
$[100]$ & $7056$ & $14976$ & $10.41$ & $10.41$ & $53.02$ \\
$[111]$ & $7200$ & $15552$ & $10.51$ & $10.92$ & $51.44$ \\
\hline \hline
\end{tabular}
\vspace{0.2cm}
\label{numpart}
\end{table}

In this study, we examine both the [100] and [111] orientations of this
NaCl crystal/binary fluid interface.
For reference, we define the $z$-axis to be perpendicular to the 
interfacial plane.  Periodic boundary conditions are applied in $x$, $y$, 
and $z$ directions.  The 
length in the $z$-direction, $L_z$, is a sum of lengths of the two separate 
blocks in the  $z$-direction and the $2 \sigma_A$-gap initially left between 
the two blocks.  The total number of particles used are 22,032 and 22,752 
for the [100] and [111] orientations, respectively. The details of the
system sizes used are summarized in Table~\ref{numpart}. 
As the systems studied are large, we have implemented the cell method 
technique~\cite{Rappaport95} to efficiently carry out the molecular-dynamics
simulations~\cite{Rappaport95}.  
Equilibration was done for about $4\times 10^5$ collisions per particle(cpp).  
During sampling, the run 
was divided into blocks of $2600$ cpp each.  
The sampling run was over a length of $30$ blocks, but since each block gives 
two independent measures of interfacial properties when the system is properly 
folded at the center of the crystal, the results reported are averages taken 
over 60 samples.

In the course of such simulations the position of the interfacial plane can shift due to
Brownian motion or due to melting or freezing of the crystal. To prevent such
motion from artificially broadening the interfacial profiles it is necessary to
monitor the position of the interfacial plane during the sampling runs. 
The standard measure of interfacial position for a planar interface
is the Gibbs dividing surface\cite{Tiller91} defined for a multicomponent system as the 
value of $z$ where the surfaces excess number of ``solvent'' particles is zero. However,
accurate calculation of the Gibbs dividing surface requires relatively long simulation
runs and is then unsuitable for a method of monitoring the time dependence of 
interfacial position. In our previous studies\cite{Davidchack98,Sibug-Aga02}, we 
find that a more suitable measure can be obtained from 
measurement of orientational order parameter profile as a function of $z$.
The orientational order parameter is defined as follows
\begin{equation}
q_n(z)=\left<\frac{1}{N_z}\sum_{i,j,k}cos\left [n\theta_{xy}(i,j,k)\right]\right>,
\label{qn}
\end{equation}
where $n=4$ for the [100] orientation and $n=6$ for the [111], $i,j$ and $k$
are nearest neighbor large particles, $\theta_{xy}(i,j,k)$ is the bond angle 
formed by $i,j$ and $k$ projected on the $x,y$ plane, and $N_z$ is the total 
number of particles that form  bond angles.  The average is taken over the number 
of angles found between $z-\Delta z/2$ and $z+\Delta z/2$, where $\Delta z$ is 
equal to the layer spacing of the bulk crystal.  

\begin{figure}[b]
\vspace{0.2cm}
\centerline{
\epsfig{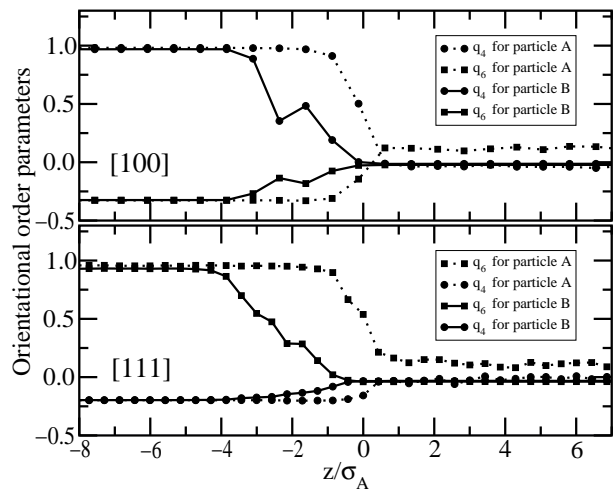}}
\caption{Large (dotted) and small (solid) particle orientational
order parameter profiles, $q_4$ (circle) and $q_6$ (square) for 
the [100] and [111] interfacial orientations.  The point $z=0$ 
is the location of the interfacial plane as calculated from the 
large particle order parameter profile.} 
\label{nord}
\vspace{0.2cm}
\end{figure}

We show in Fig.~\ref{nord} the order parameters, $q_4$ and $q_6$, of the two particle 
types for the [100] (upper panel) and [111] (lower panel) orientations. As 
expected $q_4$ ($q_6$) is small in the [111] ([100]) interface where 6-fold (4-fold) 
symmetry dominates.  We define the interfacial position relative to the
midpoint of the orientational order profile for the large particles. That is,
$z=0$ in all of the $z-$dependent profiles presented in this study is defined
as the point at which the orientational order parameter has decayed halfway from
its crystal to fluid value. This order parameter is suitable as a measure of
interfacial position since it is smoothly monotonically decreasing and can be
calculated accurately for very short runs.  The parameter 
profiles of the small particles are not smoothly varying because at the 
interfacial region, some number of small particles cluster together to occupy 
large particle vacancies at the interfacial region (as will be seen in the density 
plots presented in the next section) disrupting the smooth transition from 
crystal-like to fluid-like value of the orientational order parameter. 

Analysis of the interfacial position as a function of time shows that during the
equilibration run the crystal exhibits some initial growth, but quickly stabilizes 
before the averaging runs are begun. Brownian motion of the solid phase, as monitored by motion of
the inner layers of the crystal, was found to be negligible due to the large system size and
no correction was necessary. 

\section{Simulation Results for the [100] and [111] Interfaces}
\subsection{Structure: Density profiles and contour plots}
The structural variation of the system across the interface is determined by 
calculating the density profile for each particle type.
\begin{equation}
\rho_i(z)=\frac{<\!\!N_i(z)\!\!>}{L_xL_y\Delta z}
\end{equation}
where $i$ denotes a particle type, 
$\Delta z$ is $1/25$ of the crystal layer spacing, 
$<\!\!N_i(z)\!\!>$ 
is the average number of particles of type $i$ in the region between 
$z-\Delta z/2$ and $z+\Delta z/2$.

\begin{figure}[t]
\vspace{0.2cm}
\centerline{
\epsfig{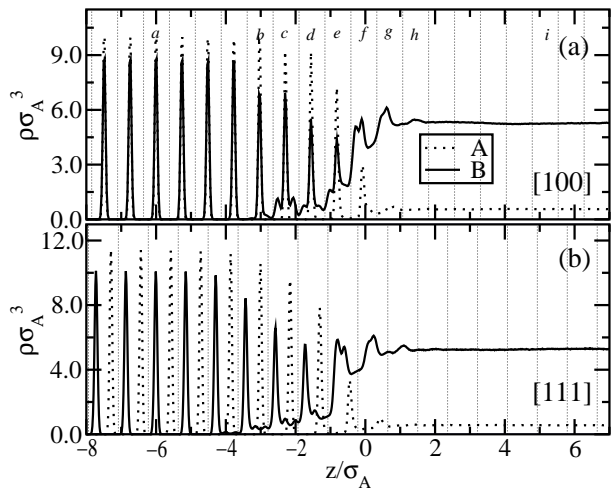}}
\caption{Large (dotted) and small (solid) particle 
density profiles of the NaCl/binary fluid interface for the 
[100] and [111] orientations.  The distance between vertical dotted 
lines is equal to the crystal layer spacing in [100] and 
twice the spacing in [111]}
\label{nfnrho}
\vspace{0.2cm}
\end{figure}

\begin{figure}[t]
\vspace{0.2cm}
\centerline{
\epsfig{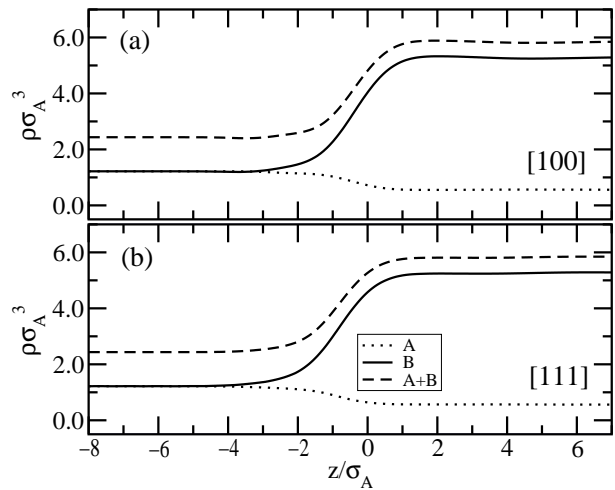}}
\caption{Filtered density profiles for the large (dotted), small (solid) and 
total (dashed) densities in the (a) [100] and (b) [111] interfacial 
orientations.}
\label{nrhofil}
\vspace{0.2cm}
\end{figure}

The density profiles of the two particle types are shown in Fig.~\ref{nfnrho}(a) 
and (b) for the [100] and [111] orientations, respectively.  
As expected for an NaCl-type lattice, the small and large particle peaks 
are in phase in [100] and exactly out of phase in [111]. This is similar to
the registry of particle density peaks found at the interfacial region of the 
lower pressure pure FCC/binary fluid system~\cite{Sibug-Aga02} .
Due to the higher pressures in this study,
the crystal peaks are much sharper than those seen in the lower pressure
binary system\cite{Sibug-Aga02} or in the single component interface\cite{Davidchack98}. 
The small side peaks in the NaCl density are due to the filling
of large particle vacancies in the lattice structure with
several smaller particles - as discussed below.  
In order to reveal any change in the lattice spacing through the interface and to index the
interfacial planes for later use, vertical dotted lines separated by the bulk crystal lattice spacing 
were added to Fig.~\ref{nfnrho}. Labels $a$ to $i$ in the [100] profile 
marks some layers whose cross-sectional
density distributions have been determined and will be discussed later.  For both orientations,
there is no discernible change in the lattice spacing as the interface is traversed from crystal to
fluid, in contrast to what was seen in our previous studies where the
crystal was either a pure or disordered FCC lattice\cite{Davidchack98,Davidchack99,Sibug-Aga02}.
In those studies there was a significant increase in the lattice spacing in the [100]
orientation as the fluid side of the interface was approached. 

The oscillations in the fine scale density profiles shown in Fig.~\ref{nfnrho} make it
difficult to see the overall trend in bulk density, so we have 
processed these profiles using a Finite Impulse Response filter~\cite{NumRec,Davidchack98}
to reveal the non-oscillatory component of the density variation.  
The resulting filtered density profiles are shown 
in Fig.~\ref{nrhofil}.  
The 10-90 width of these bulk density profiles provides a measure of the interfacial width. 
(The 10-90 width of a monotonically varying interfacial profile is the distance over which
the profile changes from 10\% to 90\% of the higher of two coexisting bulk values, relative
to the lower bulk value.) 
The 10-90 width derived from the large particle density profile of the [100] 
orientation [see Fig.~\ref{nrhofil}(a)] is $2.6\sigma_A$, corresponding to the region 
between  $z=-2.2\sigma_A$ to $z=0.4\sigma_A$.  For the small particles 
the 10-90 width is smaller at 2.3$\sigma_A$ and the 10-90 region
($z=-1.7\sigma_A$ to $z=0.6\sigma_A$) is shifted slightly toward the fluid, relative to
the 10-90 region of the large spheres.  
Combining these two regions, the interfacial region of the [100] 
orientation defined by the densities 
has a width of $2.8\sigma_A$.
The total interfacial width defined for the densities for the [111]
orientation ($2.9\sigma_A$) is not significantly different than that for
[100].  The interfacial widths of the large particle density profiles are identical to those
found in our earlier lower pressure FCC/binary fluid interface study~\cite{Sibug-Aga02} for the 
same diameter ratio, but narrower than the 3.3$\sigma$ widths found in the single component
system\cite{Davidchack98}. In contrast, the small particle interfacial widths found here are
much smaller than those found in the lower pressure study, where they were found to be about
3.3$\sigma_A$. As a consequence, the overall interface for the NaCl/fluid is slightly narrower than
the lower pressure pure FCC/fluid interface. 

\begin{figure}[t]
\centerline{
\epsfig{file=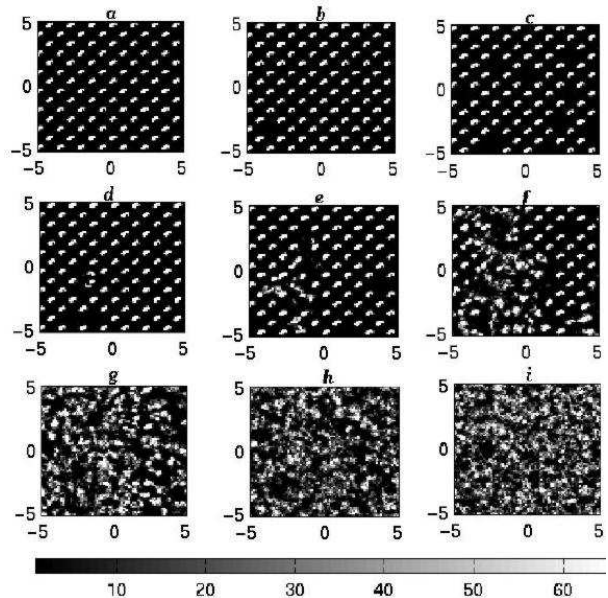,height=8.0cm,width=8.0cm,angle=0}}
\caption{Large particle density contour plots parallel to the 
interfacial plane for different layers of the [100] interface.  
The layers are as labeled in Fig.~\ref{nfnrho}(a).}
\label{n100xyrha}
\vspace{0.2cm}
\end{figure}

\begin{figure}[t]
\centerline{
\epsfig{file=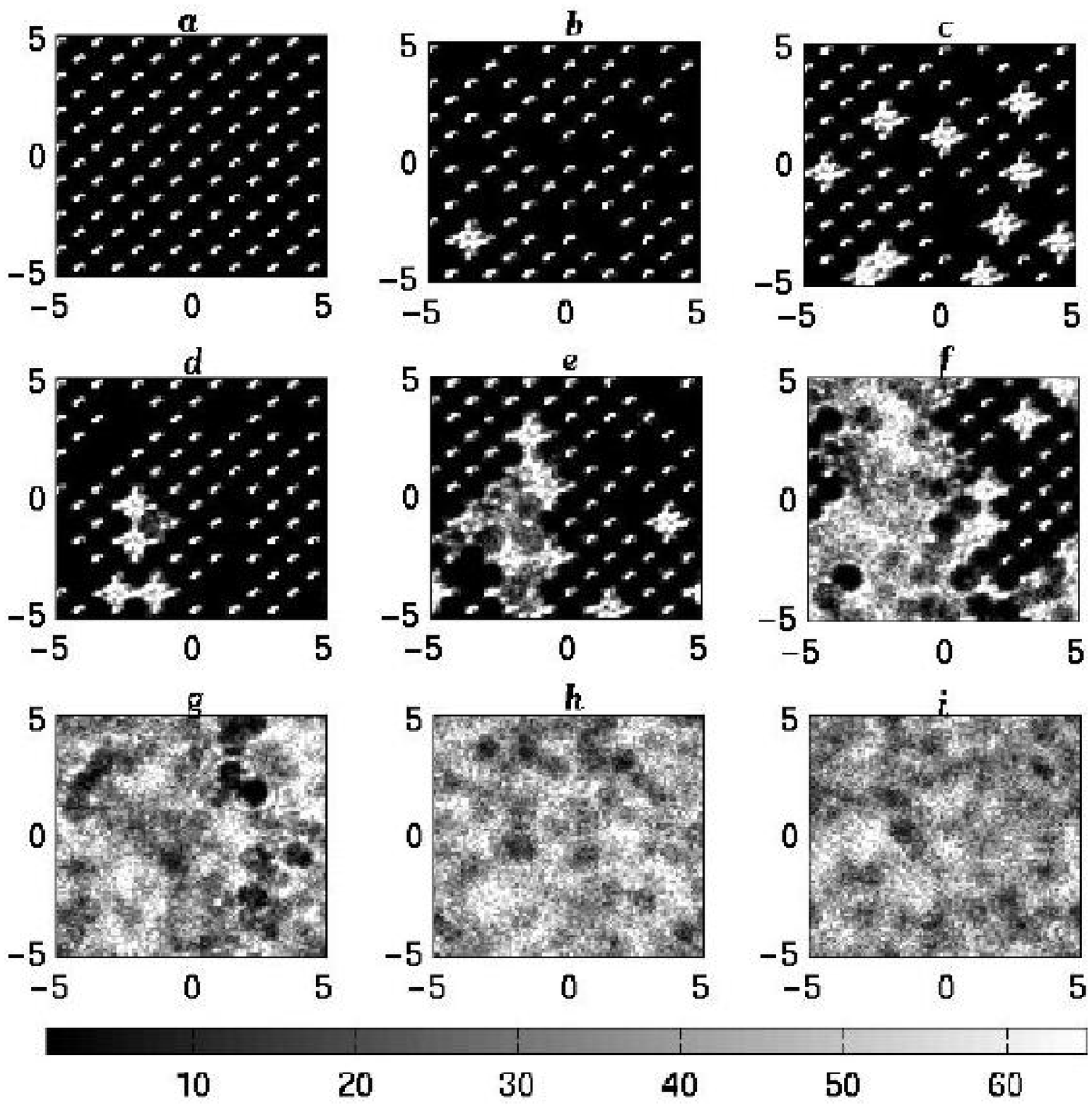,height=8.0cm,width=8.0cm,angle=0}}
\caption{Small particle density contour plots parallel to the 
interfacial plane for different layers of the [100] interface.  
The layers are as labeled in Fig.~\ref{nfnrho}(a).}
\label{n100xyrhb}
\vspace{0.2cm}
\end{figure}

\begin{figure}[b]
\vspace{0.2cm}
\centerline{
\epsfig{file=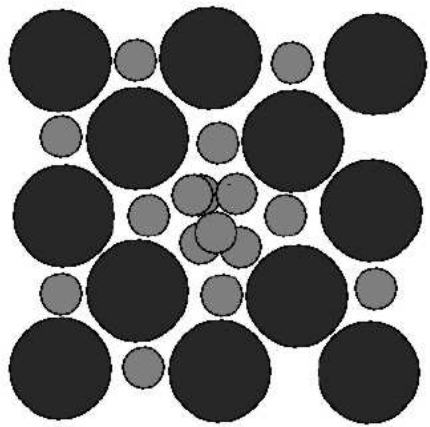,height=6.0cm,width=6.0cm,angle=0}}
\caption{Snapshot of a portion of layer $c$ as labeled in  
Fig.~\ref{nfnrho}(a) showing small particles occupying a large particle
vacancy.}
\label{snapshot}
\end{figure}

To get a detailed understanding of the structural transition 
across the interface between the two coexisting phases, we have determined the cross-sectional density
distributions within layers parallel to the interfacial plane.  For each layer, the 
2-D density distribution is defined as
\begin{equation}
\rho_i^z(x,y)=\frac{<\!\!N_i^z(x,y)\!\!>}{\Delta x\Delta y\Delta z}
\end{equation}
where $i$ denotes a particle type, $\Delta x=\Delta y=0.12 \sigma_A$, 
$\Delta z$ is the crystal layer spacing, which is $0.74 \sigma_A$ for 
[100].
The average number of particles of type $i$ in the volume given by 
$\Delta x\Delta y\Delta z$ is $<N_i^z(x,y)>$.  In Figs.~\ref{n100xyrha}  
and  ~\ref{n100xyrhb} are density contour plots of the [100] interface 
for the large and small particles, respectively.  Layers $a$ to $i$ are as 
labeled in Fig.~\ref{nfnrho}(a), where $a$ is deep into the 
bulk crystal, $i$ is in the bulk fluid and $b$ to $h$ are interfacial regions.
The decrease in density peak height of the large (Type A) spheres
in Fig.~\ref{nfnrho} starting at layer $b$ is initially due to the presence of 
lattice vacancies as shown in Fig.~\ref{n100xyrha}.  Small particle vacancies also start 
to appear in layer $b$ of Fig.~\ref{n100xyrhb}.  We also find by comparing 
Figs.~\ref{nfnrho} and ~\ref{n100xyrha} for layers $c$ and $d$ that the side 
peaks appearing at these layers are due to the accumulation of small 
particles in the large sphere vacancies.  The structure of this vacancy filling is
interesting in that the large particle is typically replaced by 6 small particles
(although a small number of vacancies filled with 5 or 7 small spheres do occur)
with little disturbance to the surrounding lattice. This can be seen in Fig.~\ref{snapshot}
where a snapshot of one of these filled vacancies in layer $c$ is shown. 
A uniform 2-D density distribution begins to develop in layer $G$ 
for both particle types indicating that, although the $z-$dependent density profiles still 
has some oscillations in this region, the structure is that of an inhomogeneous fluid at a wall.

Of particular interest to materials scientists is the degree of interfacial
segregation - the preferential adsorption (or desorption) of one component 
(usually the ``solute'') at the interface. This quantity is defined
relative to the Gibbs dividing surface.  The Gibbs dividing surface of a planar interface
is defined\cite{Tiller91} as the plane along the $z-$axis giving a vanishing
surface excess solvent particle number, $\Gamma_{\mbox{solvent}}$, defined in the equation
\begin{equation}
N^{\mbox{solvent}}/A=\rho_c^{\mbox{solvent}}z+\rho_f^{\mbox{solvent}}(L_z-z)+\Gamma_{\mbox{solvent}}
\label{gibbs}
\end{equation}  
where $N^{\mbox{solvent}}$ the total number of solvent particles spheres, $A$ is the area of the 
interface, $\rho_c^{\mbox{solvent}}$ and $\rho_L^{\mbox{solvent}}$ are the bulk densities, $z$ is the 
location of the interface assuming the length of the simulation box runs 
from $0$ to $L_z$.  Defining the small particles as the 'solvent', we find that the 
Gibbs dividing surfaces are at $z=-0.49\sigma_A$ and $z=-0.93\sigma_A$ relative 
to the position calculated from the large particle order parameter for the 
[100] and [111] orientations, respectively.  Surface excess of the 'solute' 
particles, $\Gamma^A$ was found to be negligible, indicating the absence of interfacial 
segregation, a result that is consistent for other crystal/melt systems that have 
been investigated~\cite{Davidchack98,Hoyt01,Sibug-Aga02}.

\subsection{Transport: Diffusion coefficient profiles}

Inhomogeneities in the transport properties within the interfacial region 
can be examined by calculating $z-$dependent diffusion coefficient profiles, defined  
for a particle of type $i$ by 
\begin{equation}
D_i(z)=\lim_{t \rightarrow \infty} \frac{1}{6N_i(z)}\frac{d}{dt}\sum_{j=1}^{N_i(z)}
\left<[\mbox{\boldmath$r$}_j(t)-\mbox{\boldmath$r$}_j(t_0)]^2\right>.
\end{equation}
The term in the summation is the mean-squared displacement over a time 
interval 
$t-t_0$ of a total of $N_i$ type $i$ particles located between $z-\Delta z/2$ and 
$z+\Delta z/2$ at time $t_0$, where $\Delta z$ is the layer spacing in [100] 
and is twice the layer spacing in [111].

\begin{figure}[t]
\vspace{0.2cm}
\centerline{
\epsfig{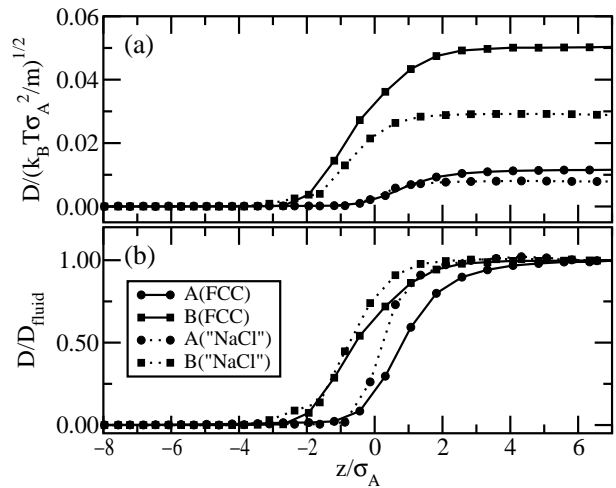}}
\caption{(a) Diffusion coefficient profiles for the [100] orientation 
of the FCC/fluid (dashed) and the AB/fluid (solid) interfaces for 
both particle types (circle for large and square for small);  
(b) Corresponding diffusion coefficient plots scaled to be zero 
in bulk crystal and unity in bulk fluid.}
\label{n100difsc}
\vspace{0.2cm}
\end{figure}

Figure ~\ref{n100difsc}(a) shows the diffusion coefficient profiles 
for the current study (dotted lines), including for comparison the 
results previously reported for the lower pressure FCC/binary fluid interface~\cite{Sibug-Aga02} 
(solid lines).  Only the [100] results are shown as the diffusion profiles for 
the [111] interfaces are not statistically different.  
The error bars are small and so are not shown for clarity 
of the plots.  The bulk fluid value for the large particles in the
lower pressure FCC/binary fluid system is $0.012 (kT\sigma_A^2/m)^{1/2}$ and that 
for the small particles is $0.050 (kT\sigma_A^2/m)^{1/2}$.  
Since the AB/fluid system has a higher pressure and larger fluid packing fraction, 
the bulk fluid diffusion coefficient values  are lower:  
$0.008 (kT\sigma_A^2/m)^{1/2}$ for the large particles and 
$0.029 (kT\sigma_A^2/m)^{1/2}$ for the small particles.  
The difference in magnitude between the small and large particle
diffusion constants makes it difficult to compare the two diffusion profiles. For
a clearer comparison in Fig.~\ref{n100difsc}(b) we plot the data
in Fig.\ref{n100difsc}(a) normalized relative to the bulk fluid values.  
Traversing the 
system from fluid to crystal, we find a region of width 
greater than $1 \sigma_A$ where the small particles have 
nonzero diffusion coefficient while the large particles have 
have effectively zero diffusion.  Both high- and low-pressure
systems exhibit this shift in the change from crystal-like to 
fluid-like motion of the two particle types. 

\begin{figure}[t]
\vspace{0.2cm}
\centerline{
\epsfig{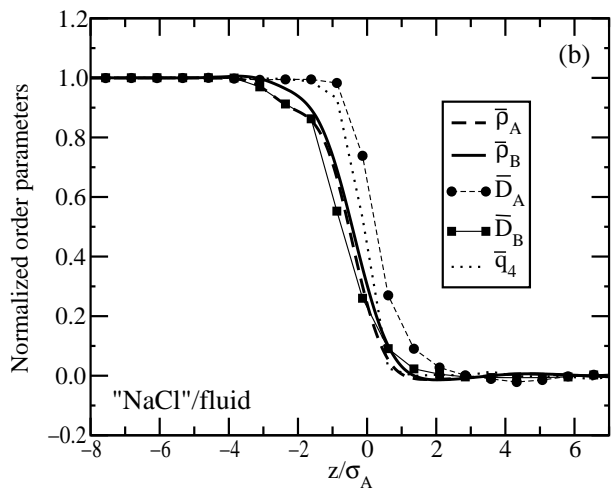}}
\caption{Filtered density, diffusion and orientational order 
parameter profiles for the [100] interface.
All profiles are scaled such that they go from unity in the crystal 
to zero in the fluid phase.}
\label{n100wid}
\vspace{0.2cm}
\end{figure}

As was done for structural transition, we can also define the extent of 
dynamical transition by determining the 10-90 region from diffusion 
coefficient profiles.  From the diffusion coefficient profile of the 
large particles this region starts from $z=-0.7\sigma_A$ up to 
$z=1.3\sigma_A$, resulting to a width equal to $2.0\sigma_A$ and centered 
at $z=0.3\sigma_A$.  The small particles define an interfacial region 
that starts from $z=-2.2\sigma_A$ and ends at 
$z=0.6\sigma_A$.  These boundaries give a width of $2.8\sigma_A$, which is 
40\% greater than the width from diffusion of the large particles. 
Also,  the midpoint is shifted by $1.1\sigma_A$ to the fluid side 
relative to the midpoint of the region defined by the large particles. 

As in our previous studies on hard-sphere interfaces\cite{Davidchack98,Sibug-Aga02}
the regions defined each of the individual density and diffusion coefficient profiles
do not all coincide, so we take a union of these 
10-90 regions to define the full interfacial region.  
Taking into account all profiles, the NaCl/binary fluid hard-sphere interface 
studied here has an overall 10-90 width of $3.5 \sigma_A$, corresponding to layers 
$c$ to $g$ in Fig.~\ref{nfnrho}.  As perhaps expected for the higher
pressure system, this interfacial region is narrower than the one found for 
lower pressure FCC/binary fluid interface\cite{Sibug-Aga02}, which was found to be $4.8 \sigma_A$.  
To summarize all of the profile data for quick inspection we show in 
Fig.~\ref{n100wid} all order parameters profiles, normalized such that we have all 
values equal to unity in the bulk crystal and zero in the bulk fluid for 
the [100] interface.  
Except for differences in the overall interfacial
width, this plot is qualitatively quite similar to the corresponding plot
for our earlier low-pressure FCC/binary fluid simulations in that 
the transition of densities for both particle types 
and the diffusion of the small particles is observed over approximately 
the same region, while the transition for the large particle diffusion is 
shifted by about $1\sigma_A$ (1.3$\sigma_A$ in the earlier study).  
Another notable feature in this plot is that 
the transition for the orientational order parameter, which we use to locate 
the interfacial plane,  occurs at about the center of these two transition 
regions. 

\section{Summary}

Using molecular-dynamics simulation, we have investigated the structure and 
dynamics of the [100] and [111] crystal-melt interfaces of the AB(NaCl)/binary 
fluid system for a two-component hard-sphere system in which the ratio of small to 
large particle diameter is $\alpha=0.414$. This system was at a pressure
of $53 kT/\sigma_A^3$ that is at the lower range of the NaCl crystal/fluid
coexistence region.  These simulations complement our earlier work\cite{Sibug-Aga02} on the
pure FCC/binary fluid interface found in this same 
system at lower pressure ($20.1 kT/\sigma_A^3$).
We find that the higher pressure AB/binary-fluid interface 
has a narrower interfacial region of $3.5\sigma_A$ compared to the 
lower pressure FCC/binary fluid system at the 
same diameter ratio, which had an interfacial region of width $4.8\sigma_A$.  
In addition, the crystal side of the higher pressure binary interface exhibited
much higher vacancy defect concentrations than either the low pressure
binary system or the single-component interface\cite{Davidchack98}. In the
interfacial region, all vacancies  in the large particle lattice were found, with
little distortion in the surrounding lattice, 
to be filled with an average of 6 small particles.
Similar to what was seen in the lower pressure FCC/binary fluid interface, the transition 
regions for both density profiles and the small particle diffusion constant are 
approximately coincident whereas the diffusion profile for the large particles
is shifted relative to the others by about $1 \sigma_A$ toward the liquid
side of the interface. 

\vspace{24pt}
\noindent
{\bf ACKNOWLEDGMENTS}
\vspace{12pt}

We gratefully acknowledge R.L. Davidchack for helpful conversations,
as well as the Kansas Center for Advanced Scientific Computing for
the use of their computer facilities. We also would like to thank the
National Science Foundation for generous support under grant CHE-9900211.

\end{document}